\documentstyle[aps,pra]{revtex}
\begin{document}
\def\la{\langle}
\def\ra{\rangle}
\def\om{\omega}
\def\Om{\Omega}
\def\vep{\varepsilon}
\def\wh{\widehat}
\newcommand{\beq}{\begin{equation}}
\newcommand{\eeq}{\end{equation}}
\newcommand{\beqa}{\begin{eqnarray}}
\newcommand{\eeqa}{\end{eqnarray}}
\newcommand{\intf}{\int_{-\infty}^\infty}
\newcommand{\into}{\int_0^\infty}
\twocolumn
\begin{title}
{\Large \bf Comment on ``Foundations of quantum mechanics: Connection with 
stochastic processes''}
\end{title}
\author{D. Alonso$^{1}$, J. G. Muga$^2$, R. Sala Mayato$^{3}$}
\address{$^1$ Departamento de F\'\i sica Fundamental y Experimental, 
Universidad de La Laguna, La Laguna, Tenerife, Spain} 
\address{$^2$ Departamento de Qu\'\i mica-F\'\i sica, Universidad del 
Pa\'\i s Vasco, Apdo. 644, 48080 Bilbao, Spain}
\address{$^3$ Departamento de F\'\i sica Fundamental II, Universidad de La 
Laguna, La Laguna, Tenerife, Spain}

\maketitle

\begin{abstract}
Recently, Olavo has proposed several derivations of
the Schr\"odinger equation from 
different sets of hypothesis (``axiomatizations'')
[Phys. Rev. A {\bf 61}, 052109 (2000)].
One of them is based on the infinitesimal inverse Weyl
transform of a 
classically evolved phase space density. 
We show however that the Schr\"odinger equation
can only be obtained in that manner for linear or
quadratic potential functions. 
\end{abstract}


The relation between classical and quantum mechanics has been 
scrutinized from many different points of view, and even so many aspects 
are still debated or unclear. 
In a recent paper \cite{Olavo00}, and in \cite{Olavo99},
Olavo proposes a derivation  
of the Schr\"odinger equation starting from 
the equation of motion of the 
normalized phase space ``density''
$F^{cl}(x,p;t)$ (we shall limit our discussion to one dimension
for simplicity),
\beq
F^{cl}_t=-\frac{p}{m}
F^{cl}_x +V_x F^{cl}_p. 
\label{Lio}
\eeq
$F^{cl}$ is associated with an ensemble of independent particles that 
evolve classically, but it may take negative values \cite{MSS93}.    
It is assumed that the force on the particle derives
from a potential $V$, and the partial derivatives with respect to
$x$, $p$ and $t$ are indicated by subscripts. The same convention
will be used later 
for other variables and second derivatives, in particular 
$A_{uv}\equiv\partial^2 A/(\partial u\partial v$).

Then Olavo defines   
a ``characteristic function''
as the inverse Weyl transform of $F^{cl}$, 
\beq
Z(x,\delta x/2;t)=\int_{-\infty}^{\infty} dp\,
F^{cl}(x,p;t) \exp(ip\delta x/\hbar). 
\label{char}
\eeq
We have introduced for later convenience a couple of notational
changes with respect to Olavo's 
work: $F^{cl}$ instead of $F$, and $Z$ instead of $Z_Q$.
The superscript ``$cl$'' will distinguish classically evolved 
quantities from quantum 
ones, labelled by ``$q$'', whenever confusion might arise.

Olavo restricts $\delta x$ to infinitesimal values, but several 
equations derived in \cite{Olavo00,Olavo99} do not require
this constraint, so we shall       
maintain by now $\delta x$ as an arbitrary   
real number. Note that we have introduced $\hbar$
in the transformation, this will allow to obtain 
equations with quantum appearance later on. 
Multiplying (\ref{Lio}) by the exponential in (\ref{char}), 
integrating over $p$, and assuming that $F^{cl}(x,p)$ vanishes for 
$p=\pm \infty$ one obtains 
\beq
i\hbar Z_t=-\frac{\hbar^2}{m}
Z_{x\delta x}
+\delta x
V_x.
\label{inter}
\eeq
With the change of variables  
\beqa
y&=&x+\frac{\delta x}{2},
\nonumber
\\
y'&=&x-\frac{\delta x}{2}, 
\eeqa
Eq. (\ref{inter}) takes the form 
\beq
i\hbar Z_t=-\frac{\hbar^2}{2m}
\left(Z_{yy}-Z_{y'y'}\right)
+(y-y')
V_{(y+y')/2} Z.
\label{exact}
\eeq
Olavo assumes that $\delta x$ is infinitesimal to discretize   
the derivative,    
\beq
(y-y')
V_{(y+y')/2}=V(y)-V(y')+O[(y+y')/2)]^2, 
\label{apro}
\eeq
so that, retaining only the first terms, 
\beq
i\hbar Z_t\approx-\frac{\hbar^2}{2m}
(Z_{yy}-Z_{y'y'})
+[V(y)-V(y')] Z.
\label{LvN}
\eeq
This equation is exact when $V$ is
linear or at most quadratic, 
since in these cases there are no correction terms in Eq. (\ref{apro});  
but otherwise it is only valid when $y$ and $y'$ are
infinitesimally close. 
It is to be noted that Eq. (\ref{LvN}) has already, 
for arbitrary $y$ and $y'$, the {\it form} of the Liouville von
Neumann quantum dynamical 
equation for the density matrix $\la y|\rho^q(t)| y'\ra$,
\beqa
i\hbar  \la y|\rho^q(t)|y'\ra_t
&=&-\frac{\hbar^2}{2m}
\left(\la y|\rho^q(t)|y'\ra_{yy}-\la y|\rho^q(t)|y'\ra_{y'y'}\right) 
\nonumber\\
&+&[V(y)-V(y')] \la y|\rho^q(t)|y'\ra.
\label{LvNq}
\eeqa
In the classical case the dynamics may be formulated similarly in
terms of a classical density matrix defined as   
\beq
\la x+{\delta x}/{2}|\rho^{cl}(t)|x-{\delta x}/{2}\ra
=Z(x,\delta x/2; t)
\eeq
for all $x$ and $\delta x$.  
In this notation Eq. (\ref{exact}) takes the form 
\beqa
i\hbar \la y|\rho^{cl}(t)|y'\ra_t
&=&-\frac{\hbar^2}{2m}
\left(\la y|\rho^{cl}(t)|y'\ra_{yy}-\la y|\rho^{cl}(t)|y'\ra_{y'y'}\right)
\nonumber\\
&+&(y-y') V_{(y+y')/2}
\la y|\rho^{cl}(t)|y'\ra.
\label{LvNcl}
\eeqa
Comparing Eqs. (\ref{LvNq}) and (\ref{LvNcl}), it is clear that 
the formal passage from classical to quantum dynamics may 
be carried out by the {\it substitution} \cite{MS92}
\beq
(y-y')
V_{(y+y')/2} \to V(y)-V(y').
\label{receipe}
\eeq
One could take this substitution as
a formal recipe to {\it construct}
the quantum evolution equation from the classical one,
but of course it does not 
represent a ``derivation'' of quantum mechanics from classical 
mechanics.    

In order to {\it derive} the Schr\"odinger equation from  
(\ref{LvN}), Olavo assumes that $\la y|\rho^{cl}(t)|y'\ra$, with 
$y$ and $y'$ infinitesimally close, factorizes as
\beq
\la y|\rho^{cl}(t)|y'\ra=\la y|\psi(t)\ra\la \psi(t)|y'\ra.
\label{fac}
\eeq
This seems a natural decomposition since one can formally separate the 
variables $y$ and $y'$ in (\ref{LvN}). 
Olavo then arrives at the Schr\"odinger equation for 
$\la y|\psi(t)\ra$ by obtaining from Eqs. (\ref{fac})
and  (\ref{inter}), in the 
small $\delta x$ limit, a  
continuity and a Hamilton-Jacobi like equation with a ``quantum
potential'' term.  
It thus might appear that quantum mechanics can be obtained simply 
by means of an infinitesimal inverse Weyl transformation from 
the classically evolved phase space density. 
However, this is not so in general. That the assumption (\ref{fac}) 
cannot hold generically may be seen already by a logical inconsistency: 
if $\la y|\psi(t)\ra$ satisfies the Schr\"odinger equation, then 
$\la y|\psi(t)\ra \la \psi(t)|y'\ra$ must satisfy the Liouville von Neumann 
equation for arbitrary values of $y$ and $y'$, which contradicts the 
fact that (\ref{LvN}) is only valid classically for infinitesimally 
close $y$ and $y'$. 
Indeed the failure of the hypothesis (\ref{fac}) for classically evolved  
density matrix may be proved explicitly. Muga and Snider have shown
that neither the positivity nor the ``purity'' of the classical 
density matrix (i.e., the factorized structure in Eq. (\ref{fac})) 
are preserved in time if these properties hold
at $t=0$ \cite{MS92}.
The case of linear or quadratic potentials is exceptional 
since the dynamical equations for 
$\rho^{cl}$ and $\rho^{q}$ become identical.    
Consequently, provided that the initial classical state of the ensemble 
is chosen with the form of Eq. (\ref{fac}) (and in such a way that
$\psi(t=0)$ may represent a valid quantum state), this form will
be preserved for arbitrary $t$ for potentials which are at most quadratic,
so the quantum density matrix 
and wave function can be obtained from the classical density matrix.    
For other potentials, the coherences (non-diagonal elements of the 
density matrices) will evolve differently because of the different ways in
which the potential acts in the classical and quantum Eqs. (\ref{exact})
and (\ref{LvNq}), even if the 
initial density matrices are equal. These
differences will be noticeable faster for larger values of $|y-y'|$. 
Eventually, the diagonal terms $\la y|\rho^{cl}|y\ra$ and 
$\la y|\rho^{q}|y\ra$ will differ too, because their evolution is not
closed, that 
is, they depend as well on the off-diagonal terms.    

Up to now, we have emphasized the density matrix formulations of 
classical and quantum mechanics. We shall now use instead the 
equivalent phase space and hydrodynamical formulations \cite{MSS93}.
This will clarify further the key role played by Olavo's
factorization assumption. The same mathematical transformation
that gives $F^{cl}$ from $\rho^{cl}$ gives also the Wigner function
$F^{q}$ from $\rho^{q}$. We shall in fact drop the superscripts
for the equations that are common to both mechanics,
\beqa
&F&(x,p;t)
\nonumber
\\
&=&\frac{1}{h}\intf d(\delta x)\,\bigg\la x-\frac{\delta x}{2}\bigg|
\rho(t)\bigg|x+\frac{\delta x}{2}\bigg\ra
e^{ip\delta x/\hbar}.
\eeqa
The evolution equation of $F^{q}$ has the form
\beq
F^q_t=-\frac{p}{m}
F^{q}_x+\frac{2}{\hbar}
\sin\left(\frac{\hbar}{2}\frac{\partial}{\partial p}
\frac{\partial}{\partial x}\right) VF^q,
\label{qLio}
\eeq
where $\partial/\partial x$ acts only on $V$. 
Multiplying Eqs. (\ref{Lio}) and (\ref{qLio}) by powers of $p$ and 
integrating over the same variable a hierarchy of ``hydrodynamical 
equations'' is obtained. The first two, which  are formally identical
for the classical and the quantum cases, are the equations of
continuity and motion,
\beqa
n_t&=&-(n\bar{p}/m)_x,
\label{conti}
\\
\bar{p}_t&=&-\frac{\bar{p}}{m}\bar{p}_x-V_x-\frac{1}{nm}(n\sigma^2)_x,
\eeqa
where 
\beqa
n&=&n(x;t)=\intf dp\,F(x,p;t), 
\\
\bar{p}&=&\bar{p}(x;t)=\frac{1}{n}\intf dp\, p F(x,p;t),
\\
\sigma^2&=&\sigma^2(x;t)=\frac{1}{n}\intf dp\, (p-\bar{p})^2 F(x,p;t).
\eeqa
The equation of motion may also be written as \cite{MSS93}
\beq
W_t=-\left[\frac{1}{2m}(W_x)^2+V+I\right],
\label{em} 
\eeq
where $I$ and $W$ are defined from their derivatives
\beqa
I_x&=&\frac{1}{nm}(n\sigma^2)_x, 
\\
W_x&=&\bar{p},
\eeqa
and by the condition  
\beq
\intf dx\,\frac{n\sigma^2}{2m}=\intf dx\,nI. 
\eeq
Eq. (\ref{em}) has the form of a modified Hamilton-Jacobi equation 
and it is valid for both mechanics. It can be shown that the
term $I$ is equal to     
the ``quantum potential'' in the case of quantum pure states \cite{MSS93}, 
\beq
Q=-\frac{\hbar^2}{2mR}R_{xx},
\eeq
where $R$ is the modulus of the pure state wave function. 
This peculiar structure has an important consequence: the hydrodynamical 
hierarchy of moment equations can be truncated exactly because the evolution
of the first moment $\bar{p}$ can be expressed in terms of the density 
$n$ and the first moment itself. Only in that case the two equations may  
be condensed in the form of a Schr\"odinger equation.
This is not the case in general either
classically or for quantum mixed states.  
The point to stress here is that if $F$ satisfies the continuity
equation (\ref{conti}) and the equation of motion (\ref{em})
(we have not specified yet whether
it is a classical or a quantum $F$), and if one assumes that  
its density matrix factorizes,  
\beqa
&F&(x,p;t)
\nonumber\\
&=&\frac{1}{h}\intf d(\delta x)\,
\bigg\la x-\frac{\delta x}{2}\bigg|\psi(t)\bigg\ra
\bigg\la \psi(t)\bigg|x+\frac{\delta x}{2}\bigg\ra
e^{ip\delta x/\hbar},
\eeqa
then it follows that $I=Q$ \cite{MSS93}, so that the hierarchy
is closed. 
In other words, Olavo's factorizability assumption leads in fact, 
within the context of a Weyl-Wigner formulation of classical or 
quantum dynamics, to the Schr\"odinger equation for pure states, 
but it does not hold for
the classical distribution functions
$F^{cl}$, with the possible exceptions already mentioned of   
linear or quadratic potentials. 
If it did, the classical dynamics for an ensemble chosen 
so that $F^{cl}(x,p;t=0)=F^{q}(x,p;t=0)$ would be equal at all 
times to its quantum counterpart. That this is not so is a
manifestation of the classical breaking of the
factorizability condition.

\acknowledgments{We acknowledge support
by The Basque Government (Proyecto ``Fundamentos de mec\'anica
cu\'antica''),
The Canadian European Research Initiative on 
Nanostructures, and The Consejer\'\i a de Educaci\'on, Cultura y 
Deportes del Gobierno de Canarias (PI2000/111)}


\end{document}